\documentclass[11pt]{amsart}
\usepackage{latexsym,amsfonts}
\usepackage{amssymb,amsmath}
\newtheorem{proposition}{Proposition}
\newtheorem{theorem}{Theorem}
\newtheorem{corollary}{Corollary}
\newtheorem{lemma}{Lemma}

\begin{document} 

\begin{LARGE}
\centerline{{\bf The Pure State Space of Quantum Mechanics}}
\vskip 3 mm 
\centerline{{\bf as Hermitian Symmetric Space.}}
\end{LARGE} 
\vskip 2 cm
\centerline{{\bf R. Cirelli, M. Gatti, A. Mani\`a}}
\vskip 0.2 cm
\centerline{{\sl Dipartimento di Fisica, Universit\`a degli Studi di Milano,
 Italy}}
\par
\centerline{{\sl Istituto Nazionale di Fisica Nucleare, Sezione di Milano, Italy}}
\par
\vskip 1.5 cm 
\centerline{{\bf Abstract}}
{\sl The pure state space of Quantum Mechanics is investigated as  Hermitian 
Symmetric K\"ahler manifold. The classical  
principles of Quantum Mechanics (Quantum Superposition Principle, 
Heisenberg Uncertainty Principle, Quantum Probability Principle) and Spectral 
Theory of observables are discussed in this non linear geometrical context.}

\par 
\vskip 1 cm
{\sl Subj. Class.:} Quantum mechanics
\par
\par
\noindent 
\vskip 1.2 cm 
\centerline{{\bf 1. Introduction}}
 \vskip 0.2 cm 

 Several models of {\em delinearization of Quantum Mechanics} have 
been proposed (see f.i. \cite{Kibble} \cite{Weinberg}, \cite{CMP1}, 
\cite{Antoniou} and \cite{Bona} for a complete list of references).
 Frequentely these proposals are supported by different motivations, but it
 appears that a common feature is that, more or less, the delinearization must
 be paid essentially by the superposition principle. 

This attitude can be understood if the delinearization program is worked out
 in the setting of a Hilbert space ${\mathcal H}$ as a ground mathematical
 structure. 

However, as is well known, the ground mathematical structure of QM is the 
manifold of (pure) states ${\bf P}({\mathcal H})$, the projective space of 
the Hilbert space ${\mathcal H}$. Since, obviously, ${\bf P}({\mathcal H})$ is
 not a linear object, the popular  way of thinking that the superposition
 principle compels the linearity of the space of states is untenable. 

The delinearization program, by itself, is not related in our opinion to attempts
 to construct a non linear extension of QM with operators which act non
 linearly on the Hilbert space ${\mathcal H}$. The true aim of the delinearization
 program is to free the mathematical foundations of QM from any reference to 
linear structure and to linear operators. It appears very gratifying to be aware of
 how naturally geometric concepts describe the more relevant aspects of ordinary
QM, suggesting that the geometric approach could be very useful also in solving 
 open problems in Quantum Theories.

Of course in ${\bf P}({\mathcal H})$ remains of the linearity are well present:
 one of our aims in this  paper is just to show that such remains are 
represented by the geodesical structure; therefore even the superposition
 principle  {\em can be delinearized} without affecting its peculiar physical
 content, suggesting moreover that manifolds of states endowed with a fair 
geodesical structure could be compatible with the superposition principle.
 Another feature we stress of our work is that also the spectral 
theory of observables has a very simple description in terms of the 
differential structure of ${\bf P}({\mathcal H})$. Indeed we will show that 
the usual linear observables are described by functions respecting geodesics 
in the technical detailed meaning coded in the definition of geolinearity. 

A very important bonus of our analysis of observables is the coming out of 
{\em suitable classes of non linear observables}; about this subject we only
 anticipate a little in this work, because it will be the content of a 
forthcoming paper \cite{Cirelli Gatti Mania}.

 Now, as is well known, ${\bf P}({\mathcal H})$ is a K\"ahler manifold 
\cite{CMP1}, but the geodesic structure of a K\"ahler manifold may be very 
involved. Therefore to look at ${\bf P}({\mathcal H})$ simply as a K\"ahler
 manifold could not be the best way to bring into focus the role of the
 geodesic structure we have stressed above. On the other hand the geodesic 
structure is particularly transparent in the 
subcategory of Hermitian symmetric spaces as one can see, in the finite
 dimensional case, in the book \cite{Kobayashi Nomizu}.  

In Sec 2 we briefly review the K\"ahler  structure of
 ${\bf  P}({\mathcal H})$. Then we discuss {\em infinite  dimensional}
 symmetric homogeneous $G$ spaces and
 their  geodesical structure. We prove  that ${\bf P}({\mathcal H})$ is a 
Hermitian symmetric $G$ space with $G=  U({\mathcal H})$. As a 
 bonus we obtain that ${\bf P}({\mathcal H})$ is simply connected,  even in 
the infinite dimensional case. 

In sec 3 we carefully discuss the Superposition Principle and show how SP is
 tied up with geodesic structure of pure state space. In Sec 4 observables are 
characterized in terms of K\"ahler structure  as $K$ functions or, 
equivalentely, in terms of geodesic structure, as geolinear functions. 

In Sec 5 we discuss Uncertainty Principle in a strong version which holds for 
 Hermitian symmetric $G$ spaces. In Sec 6 we discuss Spectral Theory and 
Quantum Probability Principle for observables, in a natural geometric way.
\par
\vskip 0.1 cm
\centerline{{\bf 2. Projective Quantum Mechanics and Hermitian $G$ spaces.}}
\vskip 0.2 cm
 Let us translate Standard Quantum Mechanics (SQM) into geometrical terms, to get
 Projective Quantum Mechanics  \cite{CLM}, \cite{CMP1}, \cite{CMP2}, \cite{CMP3}, 
\cite{CMP4}, \cite{CGM}, \cite{Goslar}, \cite{Ashtekar}, \cite{Hughston}.
\par
\vskip 0.1 cm 
 Pure states in QM are geometrically described as the points of an
infinite dimensional K\"ahler manifold ${\bf P}({\mathcal H})$, the 
{\em projective space} of the Hilbert space ${\mathcal H}$ of the system.
The points $\hat \varphi , \hat \psi $, ... of ${\bf P}({\mathcal H})$ (i.e.
 the rays of ${\mathcal H}$ generated by non-zero vectors $\varphi, \psi$ of
${\mathcal H}$) are the (pure) states of the quantum system.
 

 ${\bf P}({\mathcal H})$, as a complex  manifold, can be canonically 
regarded as a real smooth manifold with an integrable almost complex 
structure $J$. The manifold $({\bf P}({\mathcal H}), J)$ is endowed with a
 natural K\"ahlerian metric, i.e. a Riemannian metric {\sf g} such that
\par
\vskip 0.2 cm 
$1)~ {\sf g}_{\hat \varphi }(v,w)={\sf g}_{\hat \varphi} (Jv, Jw),\quad 
v,w\in T_{\hat \varphi} {\bf P}({\mathcal H}),$
\par
2) {\sl the associated fundamental $2-$form
$$
{\mathbf \omega}_{\hat \varphi}(v,w) := {\sf g}_{\hat 
\varphi}(Jv,w)
$$
\par

 is closed, hence symplectic.}
\par
\vskip 0.2 cm
The natural K\"ahler metric  of ${\bf P}({\mathcal H})$ is the 
Fubini-Study metric 
$$
{\bf g}_{\hat \varphi} (v,w)= 2 \kappa \Re({\sf v}|{\sf w}),
$$
where ${\sf v}=T_{ \hat \varphi} b_{ \varphi} (v),~{\sf w}= T_{\hat 
\varphi }b_{\varphi}(w),$  and the associated fundamental $2-$form
$$
{\mathbf \omega}_{\hat \varphi}(v,w)= 2 \kappa \Im({\sf v},|{\sf w}),
$$
where $\kappa >0$ is an (arbitrary) constant. We recall that $b_\varphi$ is the
 chart at $\hat \varphi$ \cite{CGM}. To get a correct correspondence
 with ordinary Quantum Mechanics, one must assume $\kappa = \hbar$.
\par
\vskip 0.1 cm
\par
\noindent
\vskip 0.2 cm 
{\bf Symmetric homogeneous  $G$ spaces.} Finite dimensional  homogeneous
 $G$ spaces are widely discussed in the literature. Standard reference books are 
\cite{Kobayashi Nomizu} and \cite{Helgason}. As there are only a few references 
 for the infinite dimensional setting \cite{Wilkins}, we shortly review 
definitions and properties in the context of Banach  manifolds. The proofs are
 given only in the case where the extension from finite to the infinite
 dimensional setting is not easy. By {\em ordinary  Banach manifold} 
 we   mean a second countable connected Hausdorff smooth Banach manifold $M$. 
 Let $G$ be an ordinary Banach Lie group acting on $M$. Then the pair $(M,G)$ 
will be said to be a {\em homogeneous  $G$ space} if
\par
1) the action of $G$ on $M$  is smooth and transitive;
\par  
2) the isotropy group $G_x$ at $x$ is a Lie subgroup of $G$, for $x\in M$.

Since the mapping $\phi : g \mapsto gx$ of $G$ onto $M$ is
continuous, $G_x= \phi ^{-1} (x)$ is a closed Lie subgroup of
$G$. Thus $G/G_x$ has a unique smooth (actually, analytic) structure
with the property that  $G/G_x$ is a $G$ space and the canonical
map $\pi : G \to G/G_x$ is smooth (actually, analytic) and open. To
prove that the induced surjection $\phi _x : G/G_x \to M$ is an 
 homeomorphism we have just to prove that it is open. This follows by
Theorem A.I.1 in \cite{ACM}. Arguing as in Proposition 4.3 in
\cite{Helgason} we obtain that $\varphi _x$ is a diffeomorphism.

%
%
\par
The symmetric $G$ spaces constitute an important class of 
homogeneous $G$  spaces. A {\em symmetric $G$  space} is
a triple $(M,G,s)$ where $(M,G)$ is a homogeneous $G$ space  and $s$ is
an involutive diffeomorphism  of $M$ with an isolated fixed point $o$.
 Given a symmetric $G$ space $(M,G,s)$ we construct for each
point $x$ of the quotient space $M= G/G_o$ an involutive diffeomorphism
$s_x$, called the {\em symmetry at $x$}, which has $x$ as isolated
fixed point: for $x= go$, we set $s_x = g\circ s \circ g^{-1}$. Then
$s_x$ is independent of the choice of the 
$g$ such that $x=go$. There is a unique involutive automorphism
$\sigma$ of $G$ such that $\sigma (g) = s(go)$.
\par
 In every symmetric $G$ space  $(M,G,s)$  one has 
$${\mathfrak g} ={\mathfrak h} + {\mathfrak m} $$
 where ${\mathfrak m} =\{ A\in {\mathfrak g}: \sigma _e (A)= -A \}$ (we
denote by $\sigma _e$  the derivative of $\sigma$ at $e$) 
is $Ad(G_x)$ invariant and complements ${\mathfrak h}$ in
${\mathfrak g}$.

A {\em complex symmetric $G$ space} is a complex Banach manifold $M$ which is
also a symmetric $G$ space with biholomorphic symmetries and
automorphisms. Let $M$ be a real Banach manifold. An {\em almost
complex structure J} on $M$ is a smooth tensor field on $M$ whose
value at any point  $x$ of $M$ is a complex structure $J_x$ on the
tangent space $T_x M$ at $x$. A smooth map between almost complex
manifolds is said to be {\em almost complex} if its derivative at each
point of the domain is complex linear. An almost complex structure $J$
on $M$ is said to be a {\em complex structure} on $M$ if there exists a
 smooth almost complex chart at any point $x\in M$. If $J$ is a complex
 structure on $M$, then the collection of all
such almost complex charts constitutes an atlas on $M$ whose
transition functions are holomorphic;  we can thus regard $M$ as a
complex manifold.

 Let us now come to ${\bf P}({\mathcal H})$ and give it the structure of
  infinite dimensional complex symmetric homogeneous $G$ space. We denote by
 $U({\mathcal H})$ the Banach Lie group of unitary operators of 
${\mathcal H}$, by ${\mathfrak u}({\mathcal H})$ its Lie algebra 
and by $S^1({\mathcal H})$ the unit ball of ${\mathcal H}$. The natural action
 of $U({\mathcal H})$ on $S^1({\mathcal H})$ is transitive and quotients to 
the natural action of $U({\mathcal H})$ on ${\bf P}({\mathcal H})$.

\begin{proposition} The projective space ${\bf P}({\mathcal H})$ is a 
 complex symmetric $G$ space with automorphism group 
$ G= U ({\mathcal H})$. The scalar product on 
 ${\mathfrak m}$ induces on ${\bf P}({\mathcal H})$ the Fubini-Study metric.  
\end{proposition}
 
{\em  Proof.} For
$\chi \in S^1 ({\mathcal H})$ we denote by  $U_{\hat \chi} ({\mathcal H})$ the 
stabilizer subgroup of $\hat \chi$ w.r.t. this quotient action. 
$U_{\hat \chi} ({\mathcal H})$ is a closed subgroup of $U({\mathcal H})$ 
and a  Banach Lie group with Lie algebra the subspace of 
 antiselfadjoint bounded operators commuting with the one dimensional 
proiection operator $P_\chi$ on the ray generated by $\chi$. In fact, 
$Lie( U_{\hat \chi}({\mathcal H}))$ is a splitting
subspace of ${\mathfrak u}({\mathcal H})$, so that 
$U_{\hat \chi} ({\mathcal H})$ is
a Lie subgroup of $U({\mathcal H})$. If one changes the vector $\chi$, one
obtains a  conjugate Lie group. 
 Then by standard arguments ${\bf P}({\mathcal H})$ is diffeomorphic to
the orbit space
$U({\mathcal H})/U_{\hat \chi }({\mathcal H})$ \cite{Helgason}. We also 
remark that the projection operator in 
${\mathfrak u}({\mathcal H})$ with  range  
$Lie(U_{\hat \chi }({\mathcal H}))$
$$
 A\mapsto P_\chi AP_\chi + (1-P_\chi )A(1-P_\chi )
$$
 is $Ad (U_{\hat \chi}({\mathcal H}))$ invariant. 

\vskip 0.1 cm
 For $\chi \in S^1 ({\mathcal H})$ one consider the symmetry $S$ defined 
by $S=1-2P_\chi$. Then one defines the involutive automorphism $\sigma$
of $U({\mathcal H})$  by the conjugation $A \mapsto S AS ^{-1}$. The
 stability subgroup of $\sigma$  is $U_{\hat \chi} ({\mathcal H})$. As a
consequence, the quotient space $U({\mathcal H})/U_{\hat \chi} ({\mathcal H})$
 is a symmetric $U({\mathcal H})$ space. One  could identify $(P({\mathcal H}),
 U({\mathcal H}), S)$ with $(U({\mathcal H})/U_\chi ({\mathcal H}),
 U({\mathcal H}),S)$.

To the symmetric $U({\mathcal H})$ space $({\bf P}({\mathcal H}),U ({\mathcal
 H}),S )$ it is associated the {\em symmetric Banach Lie algebra} 
$(Lie (U({\mathcal H})),  Lie (U_\chi ({\mathcal H}), \sigma )$,
where 
$$
{\mathfrak g} \simeq Lie (U({\mathcal H}))= {\mathfrak u} ({\mathcal H})
$$
$$
{\mathfrak h} \simeq Lie (U_\chi  ({\mathcal H}))= {\mathfrak u}({\mathcal H})
\cap  S ^\prime  .
$$
 i.e. the commutant of $P_\chi$ in ${\mathfrak u}({\mathcal H})$.

 Finally, ${\mathfrak m}$ is the anticommutant of $P_\chi$ in
 ${\mathfrak u}({\mathcal H})$. We remark that  $\sigma$ is an involutive norm
 preserving Lie algebra automorphism. 

We can define a $Ad(U_\chi ({\mathcal H}))$ invariant scalar product
in ${\mathfrak m}$ by

$$
(A,B):= -\hbar Tr(AB)~.
$$
 The subspace $\mathfrak{m}$ is canonically identified with the
 tangent space at $\hat \chi$. Thus we get a Riemannian metric ${\sf g}$ on
 ${\bf P}({\mathcal H})$.

The complex structure in $\chi ^\perp$ induces a $Ad(U_\chi ({\mathcal H}))$
 invariant complex structure $J$ on $\mathfrak{m}$. Thus a complex
structure is induced on ${\bf P}({\mathcal H})$. The proof that the metric 
${\bf g}$ and the complex structure $J$ correspond to the Fubini-Study metric
and to the canonical complex structure of ${\bf P}({\mathcal H})$ is a simple
adaptation of the analogous finite dimensional statement
\cite{Kobayashi Nomizu}. $\quad \Box$
 \par
\vskip 1 mm 
\begin{corollary} The projective space ${\bf P}({\mathcal H})$ is simply 
 connected.
\end{corollary}

{\em Proof.} This topological property is well known in the finite dimensional
 case \cite{Kobayashi Nomizu}. In the infinite dimensional case it was proved
 by Kuiper \cite{Kuiper} that the unitary group $ U({\mathcal H})$ 
is contractible. We have the exact sequence
$$
\pi  _0 ( U_\chi ({\mathcal H})) \to 
\pi _1 ( U({\mathcal H})) \to
 \pi _1 ( U({\mathcal H})/ U_\chi ({\mathcal H}))
\to \pi _0 ( U({\mathcal H}))~.
$$
 The exponential map 
${\mathfrak u}({\mathcal H}) \to  U({\mathcal H})$ is onto by a theorem of de la Harpe \cite{Harpe}. Therefore 
the isotropy subgroup $ U_\chi ({\mathcal H})$ is connected. Thus by 
 Proposition 1  we obtain that ${\bf P}({\mathcal H})$ is  simply connected.
$\quad \Box$  

\vskip 2 mm
{\bf The Riemannian and Hermitian case.} Let $(M,G)$ be a reductive
 homogeneous  $G$ space. We denote by $H$ its stability subgroup. So we can 
 identify $M$ with the coset space $G/H$. We denote by $o$ the equivalence 
 class of $e$.

 The {\em canonical connection} $\nabla$ on $M$ is defined by
$$
\nabla _v (Y):= [X(v),Y]_x \quad v\in T_x M
$$
 for all vector fields $Y$ defined around $x$. This definition is consistent with
 the classical definition given by Kobayashi and Nomizu \cite{Kobayashi Nomizu}.
 The canonic connection is complete.

We define, for $v\in {\mathfrak m}$ and $t\in {\bf R}$
$$
c_v :t\mapsto = \exp (tv)o~.
$$
 
 For every $v\in {\mathfrak m}$ the curve $c_v$ is a geodesic  starting from
 $o$ of the canonical connection; conversely, every geodesic from $o$ is
 of the form $c_v$ for some $v\in {\mathfrak m}$.

Torsion and curvature of the canonical connection are discussed in
\cite{ Kobayashi  Nomizu}. Every reductive homogeneous  $G$ space admits a unique torsion
free $G$ invariant affine connection, the {\em natural torsion free
connection} \cite{Kobayashi Nomizu}. The natural torsion free connection has 
the same geodesics that the canonical connection.

{\begin{proposition} 
If $(M,G,s)$ is symmetric, there is a natural bijection between the
set of all subspaces ${\mathfrak m}^\prime $ of  ${\mathfrak m}$ such
that
$$
[[ {\mathfrak m}^\prime, {\mathfrak m}^\prime ],{\mathfrak m}^\prime ]
\subset {\mathfrak m}^\prime
$$
 and the set of all complete totally geodesic submanifolds $M^\prime$
 of $M$ (through $o$).
\end{proposition}

\vskip 1 mm
%
%
%

 If $(M,G,s)$ is symmetric Riemannian, then all symmetries are isometries
 and the  canonical decomposition is orthogonal. Moreover  the canonical
 connection is the unique affine connection on $M$ which is invariant w.r.t. 
 all symmetries of $M$. We denote by $T$ its torsion and by $R$ its curvature. 

\begin{proposition}
1) $T=0$, $\nabla R=0$;
$$
R(u,v)w= -[[u,v],w] \quad {\rm for}\quad u,v,w\in {\mathfrak m}.
$$
2) For every $v\in {\mathfrak m}$ the parallel transport along $\pi
   (\exp (tv))$ agrees with the differential of the transformation
   $\exp (tv)$ on $M$.

3) for every $v\in {\mathfrak m}$
$$
\pi (\exp (tv))= \exp (tv)o
$$
 is a geodesic from $o$ and conversely, every geodesic from $o$ is of
 this form.

4) Every $G$ invariant tensor field on $M$ is parallel.

\end{proposition}
\par 
\vskip 1 mm 
 As a consequence, for every simmetric Riemannian $G$ space the
 canonical connection agrees with the natural  torsion free
 connection. Moreover, every invariant Riemannian metric on $M$, if
 any, induces the canonical connection. 
\par
\vskip 1 mm 
A {\em Hermitian symmetric $G$ space} is given by $(M,G,s,{\sf g},J)$ where
$(M,G,s,{\sf g})$ is a Riemannian symmetric $G$  space and $J$ is an almost
complex structure on $J$, which is symmetry  invariant. Then $\nabla J=0$, $J$
 is integrable, so that $(M,J)$ is a complex manifold.

\begin{proposition} Let $(M,{\sf g},J)$ with $J$ almost complex and ${\sf g}$
Hermitian metric. Then

1) if $(M,G,s,J)$ is a complex symmetric $G$ space, then $J$ is integrable
   and  ${\sf g}$ is K\"ahler.

2) if $({\sf g},J)$ is a K\"ahler structure on a  symmetric Riemann
   $G$ space $(M,G,s,{\sf g})$, then $(M,G,s,{\sf g},J)$ is Hermitian 
 symmetric $G$ space.
\end{proposition}

\vskip 1 mm 
 We refer to $(M,{\sf g},J)$ as to the K\"ahler manifold underlying
 $(M,G,s,{\sf g},J)$.
\par 
\begin{proposition} Let $(M,G,s)$ be a symmetric homogeneous $G$ space,
with isotropy group $H$. 

1) If ${\mathfrak m}$  admits some $Ad (H)$ invariant complex structure
   $I$, then $M$ admits an invariant complex structure such that the
   canonical connection is complex  and $M$ is complex  affine
   symmetric.

2) If, moreover, ${\mathfrak m}$ admits an $Ad(H)$ invariant scalar
   product which is Hermitian w.r.t. $I$, then $M$ admits an invariant
   K\"ahler metric and is Hermitian symmetric.
\end{proposition}

\vskip 1 mm  
 We recall that a connected submanifold $S$ of a Riemannian manifold $M$ 
 is {\em geodesic at $m\in S$} if, for every $v\in T_mM$, the geodesic
$c_{m,v}(t)$ determined by $v$ lies in $S$ for small values 
 of the  parameter $t$. If $S$ is  geodesic at every point of
$S$, it is called a {\em totally geodesic submanifold} of $M$.
 By the above remarks we get that geodesics and hence totally geodesic 
submanifolds of a symmetric Riemannian space can be equivalentely 
characterized as the geodesics and the totally geodesic submanifolds w.r.t. 
the canonical connection.

\par
\vskip 1 mm 
 Closed complex totally geodesic submanifolds of a  Hermitian symmetric  
$G$ space corresponds to closed complex $Ad(H)$ invariant subspaces of
 $\mathfrak{m}$ \cite{Kobayashi Nomizu}. 
 In the particular case of the projective space ${\bf P}({\mathcal H})$, 
every closed complex subspace of ${\mathfrak m}$ is $Ad(H)$ invariant. So
we have:
\par
\begin{proposition} The projective space ${\bf P}({\mathcal H})$, with the 
Fubini-Study metric, is  the K\"ahler manifold underlying to the Hermitian 
symmetric space 
$({\bf P}({\mathcal H}),  U({\mathcal H}),S, {\bf g}, J)$.
 Closed complex totally geodesic submanifolds of
 ${\bf P}({\mathcal H})$ at some point of ${\mathbf P}({\mathcal H})$
 correspond exactly to closed $J$ invariant subspaces of $\mathfrak{m}$.
\end{proposition}
\par
\vskip 1 mm

\par
 We have  shown that ${\bf P}({\mathcal H})$ is a Hermitian
 symmetric $G$ space. Hermitian symmetric $G$ spaces $M$ are naturally
 reductive  and the canonical connection, the natural torsion free
 connection and the Riemannian connection agree \cite{Kobayashi Nomizu}.
 Complete totally geodesic submanifolds through some point $m$ in $M$ 
correspond bijectively to $Ad(H)$ and $J$ invariant closed subspaces
of ${\mathfrak m}$ \cite{Kobayashi Nomizu}. In the particular case of
 $M= {\bf P}({\mathcal H})$ one sees
that $Ad(H)$ invariant subspaces of ${\mathfrak  m}$ are
precisely the $J$ invariant ones. Varying the point in  
${\mathbf P}({\mathcal H})$, we see 
that  the family of all  closed totally geodesic submanifolds of 
${\bf P}({\mathcal H})$ are identified with the family of all closed complex
 subspaces of ${\mathcal H}$. Of course, the manifold ${\bf P}({\mathcal H})$
itself is considered as totally geodesic at any point. In this way we obtain a
 geometric interpretation of  {\em Quantum Logic} ${\mathcal L}({\mathcal H})$. 

We stress that Quantum Logic has a relevant role in foundations of QM 
\cite{Mackey}. We can analogously prove that closed totally geodesic submanifolds
 of any Hermitian symmetric G space have the algebraic structure of a Quantum 
Logic. So it is very surprising and  gratifying to see that this structure
 naturally appears in the general geometrical context of Hermitian
 symmetric G spaces.
\par  
\vskip 2 mm 
\centerline{{\bf 3. Superposition principle and the geodesic structure
of ${\bf  P}({\mathcal  H})$.}}
\par
\vskip 2 mm 
 The projective space ${\bf P}({\mathcal H})$  is metrically complete; the
 distance is
$$
d(\hat \varphi , \hat \psi )= \sqrt {2\hbar}\arccos |(\varphi |\psi 
)|.
$$
\vskip 0.1 cm
The diameter of ${\bf P}({\mathcal H})$  is finite and equals 
$\sqrt {2\hbar}$. The {\em equator} of $\hat \varphi$ is the  set of all
 $\hat \psi$ such that $d(\hat \varphi , \hat \psi )=
1/2 ~diam({\bf P}({\mathcal H}))$. The {\em antipodal submanifold}
 $\hat \varphi ^\perp$ of $\hat \varphi$ is the set of
 $\hat \psi \in {\bf P}({\mathcal H})$ such that
$$
d(\hat \varphi , \hat \psi )=\sqrt {2\hbar }~.
$$ 
\par

 We remark that the antipodal submanifold $\hat \varphi^\perp $ is the maximal 
closed complex totally geodesic submanifold of ${\bf P}({\mathcal H})$
 not containing $\hat \varphi$.
\par
 Antipodality has a very remarkable  content: it
translates into ${\bf P}({\mathcal H})$ orthogonality:
\vskip 0.1 cm 
\par
a) $\hat \varphi ,\hat \psi \in {\bf P}({\mathcal H})$ antipodal
 if and only if $ (\varphi |\psi )=0,$
\par
b) $\hat \varphi ,\hat \psi \in {\bf P}({\mathcal H})$ antipodal if
 and only if $ d(\hat \varphi , \hat 
\psi ) = diam({\bf P}({\mathcal H})),$
\par
c) $\hat \varphi , \hat \psi \in {\bf P}({\mathcal H})$ antipodal if and
 only  if $\hat \psi \in 
C_{\hat \varphi},$
\par \noindent 
with $C_{\hat \varphi}$ the {\em cut locus} of $\hat \varphi$, i.e.
 the complement of the greatest open  neighborhood $U_{\hat \varphi}$ of
 $\hat\varphi$ such that any point of $U_{\hat \varphi}$ might be connected to
 $\hat \varphi$ by means of one and only one minimal geodesic. 
\par
\vskip 2 mm
 For every $\hat \varphi \in {\bf P}({\mathcal H})$, the exponential 
map
$$
Exp_{\hat \varphi} : T_{\hat \varphi} {\bf P}({\mathcal H}) \to {\bf
P}({\mathcal H})
$$
is defined on the whole $T_{\hat \varphi}{\bf P}({\mathcal H})$ and the 
{\em injectivity radius}
$$
R^i _{\hat \varphi} := sup \{ \rho >0~|~Exp_{\hat \varphi} \lceil 
B(0_{\hat \varphi}; \rho ) {\rm~is~injective~}\},
$$
($B(0_{\hat \varphi};\rho )$ is the closed ball with radius $\rho$ 
centered at $0_{\hat \varphi}$) is constant and equals $\hbar \pi$.
\par
\vskip 1 mm
The mathematical formulation of Superposition Principle in the SQM is
 very well known: (SP)
\par
{\em  With the due care to normalization properties, superpositions mean 
 ${\bf C}$-linear combinations.}
\par
\vskip 0.1 cm
Translation into ${\bf P}({\mathcal H})$ is not particularly hard: {\em
statement $SP_1$}
\par
\vskip 1 mm 
{\em for any pair of distinct points 
$\hat \varphi  , \hat \psi$ the set of all superpositions of these two states 
is ${\bf P}({\mathcal H}_{ \varphi , \psi })$, the projective of the 
two-dimensional subspace of ${\mathcal  H} $ generated by any pair
 $\varphi , \psi$ of representatives of $\hat \varphi , \hat \psi$.}
\par
\vskip 0.1 cm
 For a sharp understanding of $SP_1$ one must supply a pointed geometrical 
characterization of ${\bf P}({\mathcal H}_{\varphi , \psi})$ as a subset
 of ${\bf P}({\mathcal H})$. This is  done by looking at the geodesic
 structure of ${\bf P}({\mathcal H})$.
 Let $v\in T_{\hat \varphi}{\bf P}({\mathcal H})$, with normalized
local representative $\xi \in \varphi ^\perp$. The geodesic tangent
in $\hat \varphi$ to $v$ is 
$$
c_{\hat {\varphi},v}(t)={\bf p}(\varphi \cos \frac{t}{\sqrt{2\hbar}}
 +\xi \sin \frac{t}{\sqrt{2\hbar}} )
$$
where ${\bf p}: S^1 ({\mathcal H}) \to {\bf P}({\mathcal H})$ denotes
the canonical surjection. In particular,  
$c_{\hat {\varphi},v} (\pi \sqrt{\frac{\hbar}{2}})= {\bf p}(\xi )$, so that $\hat {\xi }$
is the (unique) antipodal point to $\hat \varphi$ lying on the
geodesic $c_{\hat {\varphi}, v}$. 
   
 More generally,  if ${\sf  v}= \rho \xi$ with $\| {\sf v}\| =\rho$
 is a local representative of $v\in T_{\hat \varphi }{\bf P}({\mathcal
 H})$, the geodesic $c_{\hat \varphi ,v}(t)$ is given by
$$
c_{\hat \varphi ,v}(t)= {\bf p}(\varphi \cos \frac{\rho t}{\sqrt{2\hbar}}
+\xi \sin \frac{\rho t}{\sqrt{2\hbar }})~.
$$

Now, using the geodesic structure of ${\bf P}({\mathcal H})$, one easily 
sees that statement $SP_1$ is equivalent to the  following
 {\em Statement $SP_2$}:
\par
\vskip 0.1 cm
 {\em for any pair
 $\hat \varphi , \hat \psi ,~(\hat \varphi \ne \hat \psi)$ the set of
all superpositions of $\hat \varphi$ and $\hat \psi$
 is the smallest totally geodesic   submanifold of ${\bf P}({\mathcal
H})$ containing $\hat \varphi , \hat \psi$.}
\par
\vskip 0.1 cm 
For a complete geometric description of the physical content of
 superposition principle, we must be able to characterize {\em single}
 superpositions of states.  We remark that $\xi$ is a representative
 vector for the
unique antipodal point to $\hat \varphi$ lying on the  geodesic
$c_{\hat \varphi , v}$.  A point $\hat \chi$ of ${\bf P}({\mathcal H})$
 belongs to $c_{\hat {\varphi}, v}$ if and  only if 
$\hat \chi \in {\bf P}({\mathcal H}_{\varphi , \xi })$ and for some
normalized vector $\varphi \in \hat \varphi$ one has
$$
\frac{(\xi  |\chi )}{(\varphi |\chi )} \in {\bf R}~.
$$
 
 In particular, if $\xi$ is a normalized representative for some
 antipodal point to $\hat \varphi$, then $\hat \chi =\widehat {\varphi
 +\xi}$ lies on  the geodesic $c_{\hat \varphi ,v}$.  Thus $\hat \chi$ lies
 on the intersection of the equator of $\hat \varphi$ with the
 geodesic $c_{\hat \varphi ,v}$. 
 Conversely, this intersection point determines the geodesic 
$c_{\hat \varphi ,v}$. Any other point of the geodesic is the  ray 
corresponding to some linear
 combination  $\alpha \varphi +\beta \psi$ with real $\alpha$ and
 $\beta$. This intersection point is convenientely characterized as 
$b_\varphi (\psi )$ (or as $ b_\psi (\varphi )$, since these rays are
equal).

 Geodesics connecting two antipodal points $\hat \varphi$ and
$\hat \psi$ describe linear combinations $\alpha \varphi +\beta \psi$
with complex quotient $\frac{\alpha}{\beta }$ and are obtained
alterating the representative normalized vector for $\hat \psi$.

 So we arrive to {\em Statement $SP_3$}:

\vskip 2 mm
{\em if $\varphi ,\psi$ are orthogonal versors and $\alpha ,\beta \in
{\bf C}-\{0\}$, and
$$
\chi = \alpha \varphi +\beta \psi~,
$$
then
$$
\hat \chi = c_{\hat \varphi , v }\left( \arctan
 ( \sqrt{2\hbar}|\frac{\beta }{\alpha} |)\right)
$$
where the tangent vector $v$ corresponds in the chart $b_\varphi$ to 
 $ e^{i\theta} \psi $, with $\theta$
 denoting the relative phase  $arg \frac{\beta}{\alpha }$ of $\alpha$ and 
$\beta$.}

\vskip 2 mm {\em Therefore the full geometric formulation of
 the {\em Quantum Superposition Principle} in ${\bf P}({\mathcal H})$
 is given by  $SP_2 +SP_3$ }.

By the above discussion we see the physical relevance of the geodesic 
structure of the manifold of states. In general, we could conclude that 
 for a purely ``kinematical" formulation of the QSP we need, as a space of states,
 a manifold $M$ equipped with a ``{\sl convenient}" geodesical structure. By
 the above discussion we conclude that such a convenient geodesical structure 
is provided by the structure of Hermitian symmetric G space. Of course, in this
 more general context, superpositions of two states are represented by the closed
 totally geodesic submanifold  they generate. 

\par
\vskip 2 mm 
 Thinking of the well known paper of Wick, Wightmann and Wigner 
\cite{WWW} we can 
define a {\em superselection sector} of a Hermitian symmetric space $M$:
\par
\vskip 0.2 cm
\noindent
 {\em a superselection sector of $M$ is a closed complex submanifold $N$
 of $M$ such that for any pair $x\in N,~y\in M-N$ there is no geodesic 
connecting  $x$ with $y$.} 
\par
\noindent
\vskip 0.2 cm 
 Thus  superselection sectors are just connected components
of $M$. Hence the projective space ${\bf P}({\mathcal H})$ does not admit any 
not trivial superselection sector. We can however introduce superselection 
sectors on Projective Quantum Mechanics by  means of disjoint union of projective spaces.
\par
\vskip 2 mm 
\centerline{{\bf 4. Quantum Superposition Principle and Observables}}
\vskip 0.2 cm

Let us remember, first of all, that {\em observables} of Projective Quantum 
Mechanics are {\em real mean value maps} of bounded selfadjoint operators on
 ${\mathcal H}$,
i.e. smooth maps 
$f:{\bf P}({\mathcal H}) \to {\bf R}$ of the type 
$f(\hat \varphi )= \langle A \rangle _{\hat \varphi}$, where for 
$A \in B_{sa}({\mathcal H})$  and
$\varphi \in {\mathcal H}$ with 
$\|\varphi \| =1$ we define $\langle A \rangle _{\hat \varphi }:= 
(A\varphi |\varphi )$.
\par
\vskip 0.1 cm
The map $\langle A \rangle$ is Hamiltonian, with Hamiltonian vector
field $v_{\langle A\rangle }$ defined by
$$
d_{\hat \varphi} \langle A\rangle (\xi )= \omega _{\hat \varphi }
(v_{\langle A \rangle }(\hat \varphi ), \xi ) \quad {\rm for}\quad \xi \in 
\varphi ^\perp ~.
$$
%

\vskip 1 mm 
A {\em Killing vector field} on a Riemannian manifold $(M, {\sf g})$ is
 a complete vector field $\xi$ whose flow preserves the Riemannian structure 
({\sl i.e.} $L_\xi {\sf g}= 0$). The following theorem was proved in
  \cite{CGM}. 
\par
\vskip 0.1 cm 
 \begin{theorem}
 A vector field $\xi$ on ${\bf P}({\mathcal H})$ is Killing if
 and only if there  is a selfadjoint operator $A\in L(H)$ such that $\xi = 
v_{\langle A\rangle }$.
\end{theorem}
\par
\vskip 0.1 cm
  A Hamiltonian function $f$ on a K\"ahler manifold  
is said to be a {\em $K$ function} if its Hamiltonian vector field $v _f$ 
is Killing. 

 A smooth map $f: {\bf P}({\mathcal H}) \to {\bf R}$ 
is {\em  geolinear} if
$$
f(c_{\hat \varphi , v}(t))=
 f(\hat \varphi ) + (\sin \frac{t}{\sqrt{2\hbar}} \cos 
\frac{t}{\sqrt{2 \hbar}} )d_{\hat \varphi } f(v) +
 \sin ^2 \frac{t}{\sqrt{2\hbar}} Hess_{\hat \varphi } f(v,v), 
$$
$(\hat \varphi \in {\bf P}({\mathcal H}), v\in T_{\hat \varphi }
{\bf P}({\mathcal H})$ is a versor and $ t\in {\bf 
R})$, where $c_{\hat \varphi , v}$ is the geodesic through $\hat \varphi$ 
along $v$. We have proved in \cite{CGM}: 
\par
\noindent
\vskip 0.1 cm 
\begin{theorem} A map $f:{\bf P}({\mathcal H}) \to {\bf R}$ is geolinear if 
and  only if there is a selfadjoint operator $A\in L({\mathcal H})$ such that 
$f= \langle A\rangle$.
\end{theorem}

 Thus the $K$ functions on ${\bf P}({\mathcal H})$ are precisely the 
geolinear maps  and can be characterized as functions 
preserving (in this  particular sense) the superpositions and correspond to 
expectation value functions of bounded self adjoint operators on 
 ${\mathcal H}$.
 
We see from thms 1,2 that in Projective Quantum Mechanics
 there is a strict link, as expected, 
between observables and the dynamical vector fields; but the really 
remarkable feature is the characterization of observables (as {\sl 
geolinear maps}) and of dynamical vector fields (as {\sl Killing 
vector fields}) with no more reference to mean value maps.
\par
\par
\vskip 2 mm
  It could be of interest also to consider some  Hamiltonian
 dynamics which  not necessarily respect the Riemannian structure. 
 So also non linear observables and non linear dynamic evolutions could  
  be suitably introduced. We will discuss this important point in a fortcoming
 paper \cite{Cirelli Gatti Mania}. Here we simply want to anticipate, as an
 example, some  of such {\em flexible observables} 
 on ${\bf P}({\mathcal H})$. 

We consider as example of a  flexible observable the function 
$$
F:= \langle A\rangle \langle B \rangle \quad A,B\in B_{sa}({\mathcal H})~.
$$
Since
$$
(d_{\hat \varphi} \langle F\rangle ) ( w)=
 \langle B\rangle _{\hat \varphi} \omega ({ v}_A (\hat\varphi ) , {w}) + 
\langle A\rangle _{\hat \varphi }\omega ({ v}_B (\hat \varphi ) , { w})
$$
we have
$$
v_{\langle F\rangle } = \langle A\rangle v_B + \langle B\rangle v_A~.
$$

\par
\vskip 0.2 cm
 We stress that by the above discussion we can conclude that 

\par
\vskip 0.1 cm
{\em in principle it is possible to maintain the QSP in a non linear QM 
 provided the following conditions are respected:
\par
\vskip 0.1 cm 
i) the space of pure states is a symmetric Hermitian  
manifold $(M, G,s, J, {\sf g})$,
\par
ii) the superpositions of $x,y \in M, ~(x\ne y)$, are the points of 
the smallest  closed $J$ invariant totally geodesic submanifold
containing $x$ and $y$,
\par
iii) the observables are those maps $f: M \to {\bf R}$ that preserve 
superpositions (the $K$ functions),
\par
iv) the dynamical evolution is given by a vector field that preserves 
the Riemannian structure, i.e. by a Killing vector field on
 $(M, {\sf g})$.}

 We could add:
\par

${\rm iv}^\prime {\rm)}$ {\em the flexible dynamical evolution is given by Hamiltonian vector fields associated to some selected family of flexible
 observables}

So we can introduce  non linear dynamics.   

\par
\vskip 2 mm

\centerline{{\bf 5.  Uncertainty Principle and Hermitian structure.}}
\par
\vskip 1 mm  In SQM for each observable   $A\in B_{sa}({\mathcal H})$ the
 {\sl dispersion} in the ``state" $\varphi$ is introduced;
$$
\Delta _\varphi A := \| A\varphi -(\varphi |A\varphi )\varphi )\|, 
\quad \varphi \in S^1 ({\mathcal H}),
$$
 and the Heisenberg Uncertainty Principle  (HUP) is stated:
\vskip 0.1 cm
\begin{proposition}
 For every $A,B\in B_{sa}({\mathcal H})$ and every $\varphi \in 
S^1 ({\mathcal  H})$ the Heisenberg Inequality holds:
$$
\Delta _\varphi A.\Delta _\varphi B \ge 1/2  |(\varphi | 
[A,B]\varphi )|~.
$$
\end{proposition}

Therefore, since 
$$
\{ \langle A\rangle, \langle B\rangle \} = \langle -i/\hbar  [A,B]\rangle ,
$$
 and 
$$
\Delta _\varphi A = \sqrt {\hbar  /2} \| v _{\langle 
A\rangle } (\hat \varphi ) \| _{\mathbf g},
$$
the Heisenberg Inequality can be written
$$
|\{ \langle A \rangle , \langle B\rangle \} (\hat \varphi )|
\le \| v _{\langle A\rangle} (\hat \varphi )\| _{\mathbf g}
\| v_{\langle B \rangle } (\hat \varphi )\| _{\mathbf g}~,
$$
that is
$$
|{\mathbf \omega }_{\hat \varphi} (v _{\langle A\rangle} (\hat 
\varphi ), v _{\langle B\rangle }(\hat \varphi ))|
\le
  \| v _{\langle A \rangle } (\hat \varphi )\| _{\mathbf 
g} \| v _{\langle B\rangle } (\hat \varphi )\| _{\mathbf g}.
$$
\vskip 0.2 cm
 Heisenberg Inequality is nothing more than the uniform  
continuity of the symplectic form (or of the Poisson product) with 
respect to the topology induced on the tangent space by the Riemannian 
structure. The above argument also works for any pair of smooth functions, so
 we are lead outside of the realm of ordinary QM.

\vskip 0.1 cm
{\em Rebus sic stantibus} we can say that the HUP can be formulated in 
a general setting which does not depend on the linearity properties of 
the setting.
\par
\vskip 0.1 cm
 Let $M$ be a manifold endowed with a symplectic 
structure ${\mathbf \omega}$ and a metric structure ${\sf g}$. We 
say that the HUP holds in $(M,{\mathbf \omega}, {\sf g})$ if the 
symplectic form is uniformly continuous with respect to the topology 
of the tangent space induced by the metric, i.e. if:
$$
\exists a \in {\bf R}_+~{\rm such~that,~}\forall x \in M,
 \quad |\{ f,h \} (x)|\le a \| v _f (x) \| _{\sf g} \| v _h (x) \| 
_{\sf g}\quad \quad {\rm(HUP)}
$$
 for any pair of Hamiltonian functions $f,h$ (with Hamiltonian vector field 
$v _f$ and $v _h$, respectively).  Indeed, if HUP holds and we define
$$
\Delta _x f := \sqrt{ r /2} \| v _f (x) \| _{\sf g}
$$
 where
$$
r:= {\rm min} \{ a\in {\bf R}_+ {\rm ~such~ that~HUP~holds} \}
$$
then
$$
\Delta _x f \Delta _x h \ge 1 /2 |\{ f,h\} (x)|.
$$

 One could also introduce the dispersion function for a field $X$ by
$$
\Delta _x X:= \sqrt{r/2} ({\sf g}_x (X_x ,X_x ))^{1/2}
$$
getting 
$$
\Delta _x X \Delta _x Y \ge r/2 |\omega _x (X_x ,Y_x )|~.
$$

Therefore, adding a fifth requirement to the requirements i) to iv) 
above, we can draw the conclusion.
\par
\vskip 0.2 cm
\noindent
 {\em In principle it is possible to maintain the QSP and the HUP in a 
non linear quantum mechanics assuming i) to iv) as above}
\par
\noindent
\vskip 0.1 cm 
and
\par
\vskip 0.1 cm 
\noindent
{\em  v) the Riemannian manifold $(M, {\sf g})$ of pure states is 
endowed with a symplectic form ${\mathbf \omega}$ which is 
 uniformly continuous with respect to ${\sf g}$.}
\par
\vskip 0.1 cm 
 In particular, this holds in any Hermitian symmetric space $M$, but we can say
 more. Actually, HUP holds with $r=1$ since the complex structure operator is
 unitary. Moreover  the set of  Killing vector fields of $M$ is {\em full},
 i.e.  every tangent vector $v$ at $x$ belongs to some Killing vector field, 
for every  $x\in M$. To see this, given $y\in M$, choose some $g\in G$ such 
that  $y=gx$. Then  define $\xi _y := g_* (v)$. One easily verifies that $\xi$
 is a (well defined) Killing vector field.

 In any Hermitian symmetric space $M$, HUP can be stated in a strong form.

\begin{proposition} If $M$ is a Hermitian symmetric space, then for every
 vector field $X$ and every $x\in M$ there exists a Killing vector 
field $K$ such that
$$
\Delta _x X  \Delta _x K = |\omega _x (X_x , K_x )| ~.
$$
\end{proposition}

{\em Proof}. We can assume $\Delta _x X \ne 0$. We know by HUP that for every
 vector field $Y$ there exists $\lambda \in {\bf R}_+$ such that
$$
\lambda ^2 {\sf g}_x (Y_x ,Y_x) \ge (\omega _x (X_x , Y_x ))^2~.
$$
In particular, this holds for any Killing vector field $K$ such that
$$
K_x = Y_x =J_x X_x ~.
$$
We obtain
$$
\lambda ^2 {\sf g}_x (K_x , K_x ) = \lambda ^2 {\sf g}_x (X_x ,X_x ) \ge 
(\omega _x (X_x , J_x X_x ))^2 =( {\sf g}_x  (X_x , X_x ))^2
$$
so that 
$$
\lambda ^2 \ge {\sf g}_x (X_x , X_x ) = 2 \Delta ^2  _x X
$$
 as required. $\quad \Box$ 
\par
\vskip 1 mm 
 \vskip 1 mm 
 A natural physical requirement is that all Killing vector  fields are 
Hamiltonian. This is  true if $M$ is simply connected. In  every simply
 connected Hermitian symmetric space $M$, the set of $K$
 functions is {\em full}, i.e. the differentials of $K$ functions span the 
whole cotangent space $T_x ^* M$, for every $x\in M$. This implies that for 
every $K$ function $f$ and $x\in M$ there exists a $K$ function $h$ such that
$$
|\{ f,h\} (x)| = \| v_f (x)\| _{\sf g}\| v_h (x)\| _{\sf g}
$$
 (see Proposition 4.5 in \cite{CMP1}).   
\par 
The Cartan Ambrose Hicks Theorem allows one to characterize simply connected 
symmetric (complex) manifolds as (complex) Banach manifolds admitting a
 geodesically complete torsion free  affine connection whose curvature tensor
 is parallel \cite{Wilkins}.   
\par
\vskip 1 mm 
The dispersion function is well defined also for non geolinear functions. 
 In particular,  for a flexible observable we have
$$
\Delta ^2 _{\hat \varphi} (\langle F\rangle ) = 
\langle B\rangle ^2 _{\hat \varphi } 
\Delta _{\hat \varphi }^2 (\langle A \rangle ) +
\langle A \rangle ^2 _{\hat \varphi}\Delta ^2 _{\hat \varphi } (B) +
 2\langle A \rangle _{\hat \varphi }\langle B\rangle _{\hat \varphi } 
\langle A\circ B \rangle _{\hat \varphi}~,
$$  
 where $A\circ B$ denotes the Jordan product. In particular, if $A$ and $B$
 commute, 
$$
\Delta ^2 _{\hat \varphi} (\langle F\rangle ) = 
\langle B\rangle ^2 _{\hat \varphi } 
\Delta _{\hat \varphi }^2 (\langle A \rangle ) +
\langle A \rangle ^2 _{\hat \varphi}\Delta ^2 _{\hat \varphi } (B) +
2 \langle A \rangle _{\hat \varphi }\langle B\rangle _{\hat \varphi } 
\langle AB \rangle _{\hat \varphi}~.
$$  
 \par
\vskip 2 mm
\centerline{{\bf 6.   Spectral Theory and Quantum Probability Principle. }}

In this section we show that ordinary spectral theory for self adjoint operators 
can be easy recovered by  the corresponding Killing vector fields and the
 dispersion function. But it has to be stressed that this formulation works very
 well also for
 {\em non Killing} vector fields. This opens the possibility to found a non linear
 spectral theory. For previus attempts in this direction, see \cite{Ashtekar}.

\vskip 2 mm 
 We can define  the spectrum of $\langle A \rangle$ for $A\in
B_{sa} ({\mathcal H})$. We say that
\par
\vskip 2 mm 
a) $\lambda \in {\bf R}$ is a {\em  regular value} if 
$$
\exists \epsilon >0~|~ \langle (A-\lambda )^2 \rangle _{\hat \varphi }
>\epsilon \quad 
 \forall \hat \varphi \in {\bf P}({\mathcal H})~,
$$
\par
b) $\lambda$ is a {\em eigenvalue} if
$$
 \langle (A-\lambda )^2 \rangle _{\hat \varphi }=0 
\quad {\rm for\quad  some}\quad \hat \varphi \in {\bf P}({\mathcal H})~,
$$
\par
c) $\lambda$ {\em belongs to continuous spectrum} if
$$
\langle (A-\lambda )^2 \rangle _{\hat \varphi }>0 ~\forall \hat
\varphi \in {\bf P}({\mathcal H}) \quad {\rm and}\quad
\exists \{\hat \varphi _n \} \quad {\rm such~that}
\quad \langle (A-\lambda )^2 \rangle _{\hat \varphi _n} \to 0~.
$$

 This definition of spectrum agrees with the standard 
one for $A\in B_{sa}({\mathcal H})$.
 However, this definition immediately extends to {\em every Hamiltonian
function} on ${\bf P}({\mathcal H})$.
\par
\vskip 2 mm
 Now we  discuss spectral aspects in terms of the Hamiltonian
vector field $v_{\langle A \rangle }$. 

First, we remark that
$$
v_{\langle (A-\lambda )\rangle }= v_{\langle A \rangle}
$$
so that the  Hamiltonian vector fields, alone, do not allows to
 characterize the
 spectral points. However, since
$$
\hbar \| {\sf v}_{\langle A\rangle } (0)\| = \| A\varphi - (A\varphi
|\varphi )\varphi \|
$$
we get that $v_{\langle A\rangle} (\hat \varphi )=0$ if and only if
$$
A\varphi = \lambda \varphi \quad {\rm with} \quad \lambda = (
A\varphi |\varphi )~.
$$
 To get eigenvectors of $A$ consider those versors $\varphi$ such that $v
 _{\langle A \rangle} (\hat \varphi )=0$; the  corresponding eigenvalue
 is given by $\langle A\rangle _{\hat \varphi }$.

We can also characterize the points of continuous spectrum. A 
$\lambda \in {\bf R}$ belongs to the spectrum of $A$ if and only if
for every $\epsilon >0$ there exists a versor $\varphi$ such that
$$
\| (A-\lambda )\varphi \| <\epsilon.
$$
 This implies that there exists a sequence $\{ \varphi _n\} $ of
 versors  such that
$$
\lim _n (A\varphi _n |\varphi _n ) = \lambda
$$
so that
$$
\| (A-(A\varphi _n |\varphi _n ))\varphi _n \| \to 0~.
$$
This means that the sequence  $\{ (A\varphi _n |\varphi _n )\}$ is
Cauchy and that $\Delta _{\varphi _n } A \to 0$. The last condition amounts 
to the request that the local expression of the field 
$v_{\langle A \rangle } (\hat \varphi _n )$ in the chart $b_{\varphi _n}$ 
goes to 0 for $n \to \infty$. 
 
Conversely, let  $\lambda \in {\bf R}$ such that for some sequence
 $\{\hat \varphi _n \}$

$$
\langle A \rangle _{\hat \varphi _n} \to \lambda \quad {\rm and}\quad
\Delta _{\varphi _n } A \to 0
$$
 then $\lambda$ belongs to the spectrum of $A$; if, morever, it does
 not exists any $\hat \varphi$ such that
$$
\langle A \rangle _{\hat \varphi } = \lambda \quad {\rm and} \quad
v_{\langle A \rangle } (\hat\varphi ) =0
$$
 then $\lambda$ belongs to the continuous spectrum of $A$.

We also remark that 
$$
\hbar \| {\sf v} _{\langle A \rangle } (0) \| ^2 = \| A\varphi
-(\varphi |A\varphi ) \varphi \| ^2
=
\langle A ^2 \rangle _{\hat \varphi } - \langle A \rangle ^2 _{\hat \varphi } 
$$
 We conclude that $\lambda$  belongs to the spectrum of $A$ if and only if
  there exists a sequence $\{ \hat \varphi _n \}$ such that
$$
\langle A \rangle _{\hat \varphi _n } \to \lambda \quad {\rm and} \quad
\langle A ^2 \rangle _{\hat \varphi _n }\to \lambda ^2~.
$$
 \par 
\vskip 0.2 cm
 For a given linear operator $A$ defined on ${\mathcal H}$, $A\ne {\bf 0}$, 
 we define the {\em regularity domain} of $A$ as the open set
$$
{\bf P}({\mathcal H}) -{\bf P}(Ker A) =\{ \hat \varphi \in {\bf
P}({\mathcal H})~|~ A\varphi \ne 0 \}~.
$$
 
 We observe that $A$ quotients to a transformation 
$$
\hat A : {\mathcal D}_{\hat A} \to {\bf P}({\mathcal H})
\quad \hat A \hat \varphi := \widehat {A\varphi}
$$ 
 where ${\mathcal D}_{\hat A}:= {\bf P}({\mathcal H})- {\bf P}(Ker
 A)~.$ The transformation $\hat A$ is smooth if and only if 
$A\in B({\mathcal H})$. 
\vskip 2 mm 
 We can use regularity domains to characterize the spectra of
bounded selfadjoint operators. Let $A\in B_{sa}({\mathcal H})$. Then
$\lambda \in {\bf R}$ is said to be  {\em regular value for $A$} if
\par

\centerline{
1) $\quad {\mathcal D}_{\widehat{ (A-\lambda)}}= {\bf P}({\mathcal H})$ and
2) $\quad \widehat {(A-\lambda)}$ is  a diffeomorphism.}
\par
\vskip 2 mm 
  We say that   $\lambda$ is a {\em spectral value for $A$} if it is not a regular value.
 \par
\vskip 2 mm
This means that or
$$
(1) \quad {\mathcal D}_{\widehat {(A-\lambda )}}\ne {\bf P}({\mathcal H})
$$
or 
\par

\centerline{$ (2) \quad {\mathcal D}_{\widehat {(A-\lambda )}}=
{\bf P}({\mathcal H})\quad $ but $\quad \widehat {(A-\lambda )}\quad $ 
is not a diffeomorphism.}
\par
\vskip 1 mm 
 We remark that
$\widehat{(A-\lambda )}$ is a smooth bijection, but its inverse can
fail to be smooth. Spectral values $\lambda$ of type (1) are said to
be {\em eigenvalues of $\hat A$}. Spectral values of type (2) are said
to belong to the {\em continuous spectrum of $\hat A$}. This
definition of spectrum agrees with the precedent one.

\par
\vskip 2 mm
{\bf Remark.} We know that $(A-\lambda )^{-1}$, if it exists, is a
linear operator defined on all the Hilbert space ${\mathcal H}$,
therefore the quotient map $\widehat {(A-\lambda )^{-1}}$ exists and
agrees with $\widehat {(A-\lambda )}^{-1}$; hence  
$\widehat {(A-\lambda )}^{-1}$ is smooth  if and only if $(A-\lambda
)^{-1} \in B({\mathcal H})$.

\par
\vskip 2 mm 
 Coming to the probabilistic interpretation, we remember that in SQM the following rule is 
 posited. 
\par

\vskip 0.1 cm
{\em  The probability that a measurement of the observable $A$
 (s.a. operator of ${\mathcal H}$) in the state $W$ (von Neumann density
 operator on ${\mathcal H}$)
 gives an outcome in a Borel set $X$ on ${\bf R}$ is given by
$$
P(A,W,X)= Tr (WQ^A (X)),
$$
where
\par

\centerline{$Q^A$ is the spectral measure of $A$.}}
\vskip 0.1 cm
\par
 A {\em quantum probability 
measure} for ${\mathcal H}$ is a map $\mu : {\mathcal L}({\mathcal H})
\to {\bf R}_+$ such that $\mu ({\bf 1})=1$ and 
$$
\mu (P+Q)= \mu (P)+ \mu (Q)
$$
 whenever $P$ and $Q$ are orthogonal. Remember that 
${\mathcal L}({\mathcal H})$ denotes the  Quantum Logic of ${\mathcal H}$. 
 
If, whenever $\{ P_i \}_{i\in I}$ is a family of mutually orthogonal
projections, $\sum _i \mu (P_i )$ is convergent and
$$
\mu (\sum _i P_i ) = \sum _i \mu (P_i )~,
$$
then $\mu$ is said to be {\em completely additive}. Completely
additive quantum probability measures form a convex set, {\em the set of states
 of ${\mathcal H}$}.

To every ray $\hat \varphi$ we can associate the projection operator 
on the ray $P_{\hat \varphi}$. Then the map 
$$
\mu _{\hat \varphi} : {\mathcal L}({\mathcal H}) \to {\bf R}
 \quad Q \mapsto \mu _{\hat \varphi} (Q):= Tr (P_{\hat \varphi} Q)
$$  
 is a completely additive quantum probability mesure. Moreover, $\mu _{\hat
 \varphi}$ is pure, i.e. cannot be not trivially expressed as convex
 combination of quantum probability measures.

The essential content of Gleason Theorem is that every 
completely additive quantum probability 
measure $\mu$ on ${\mathcal L}({\mathcal H})$
has a unique extension to a positive normal functional $\Phi _\mu$  on
$B({\mathcal H})$, whenever $dim ({\mathcal H}) >2$.
 This implies that there exists a unique positive, selfadjoint trace
class operator ({\sl density operator}) $W$ such that $Tr W=1$ and  
$$
\Phi _\mu (A) = Tr (WA) \quad \forall A\in B({\mathcal H})~.
$$
 Equivalentely,
$$
\mu (Q)= Tr (WQ)
$$
for every projection operator $Q$. In particular $\Phi _\mu$ is pure
if and only if $W$ is a one dimensional projection operator, i.e. if
$\mu =\mu _{\hat \varphi}$ for some $\hat \varphi \in {\bf
P}({\mathcal H})$.

\vskip 1 mm 
Therefore, mixed states can be interpreted as probability measures on
the family of closed totally geodesic submanifolds of 
${\bf P}({\mathcal H})$. Elements of ${\bf P}({\mathcal H})$ (pure
states)
 corresponds
precisely to pure probability measures. 

 It is well known by spectral theory of density operators that
 functionals $\Phi _\mu$ can be uniquely expressed as probability
 measures on ${\bf P}({\mathcal H})$, or else as $\sigma$ convex
 combinations of pure states. 

\vskip 2 mm
 We are able to characterize the trace
functional  and probability transition map 
in terms of the metric structure of ${\mathbf P}({\mathcal H})$.
 In the projective space totally geodesic submanifolds $M$ correspond to 
projection operators $Q_M$ on the Hilbert space ${\mathcal H}$, with the 
property that $M$ is  canonically identified with the projective space of the
 range of $Q_M$. Then $\langle Q_M \rangle$ is the unique geolinear map such that 
$$
\langle Q_M \rangle _{\hat \varphi }=1~, ~ \forall \hat \varphi \in M~,
\quad
\langle Q_M \rangle _{\hat \varphi }= 0~, ~ \forall \hat \varphi \in M^\perp
 ={\bf P}(Ker Q_M)~.
$$
 In particular, there is a unique geolinear map 
$\langle Q _{\hat \varphi} \rangle $ such that
$$
\langle Q_{\hat \varphi} \rangle _{\hat \varphi }=1
\quad
\langle Q_{\hat \varphi}\rangle _{\hat \psi } =0 \quad {\rm for}\quad \hat \psi \in \hat\varphi ^\perp~.
$$
  So we obtain the probability transition map 
$$
\langle \hat \varphi |\hat \psi \rangle :=
 \langle Q_{\hat \varphi}\rangle _{\hat \psi }~.
$$
\par
\vskip 2 mm 
So traces and probability transitions are obtained as the corresponding $K$
 functions: for $\hat \varphi\in {\bf P}({\mathcal H})$
$$
\langle Q_M\rangle _{\hat \varphi }= Tr(Q_MP_{\hat \varphi}) =
 \sqrt{2\hbar} \inf_{\hat \psi \in M}\arccos|(\varphi|\psi )|=
 d(\hat \varphi , M)~.
$$

 Therefore, for a pure state $\mu =\mu_{\hat \varphi}$ we have
$$
P(A,P_{\hat \varphi},X)=\mu _{\hat \varphi}(Q^A(X))= d(\hat \varphi, M^A (X))
$$
 where $M^A (X)$ is the totally geodesic submanifold of 
${\bf P}({\mathcal H})$ canonically associated to the projective space of the 
range of the projection operator $Q^A (X)$, for a given Borel set $X$. 
\par
 The submanifold $M^{ A }(X)$ can be characterized as the 
unique closed totally geodesic submanifold of ${\bf P}({\mathcal H})$ such
 that
$$
\langle A \rangle _{\hat \varphi} \in X \quad {\rm for} \quad 
\hat \varphi \in M^{ A } (X)
$$
$$
\langle A \rangle _{\hat \varphi } \in {\bf R}-X \quad {\rm for} \quad
\hat \varphi \in \left( M^{ A  } (X)\right) ^\perp~.
$$

Every mixed state $\Phi _\mu$, associated to some density operator $W$ is a 
$\sigma$ convex combination of pure states $\Phi _\mu = \sum _i \alpha _i 
\Phi _{\mu _i}$, with $\mu _i = \mu _{\hat \varphi _i}$. Therefore we obtain
$$
P(A,W,X)= \sum _i \alpha _i d(\hat \varphi _i , M^A(X))~.  
$$    
This gives the geometric content of the probabilistic interpretation. 
\par
\vskip 2 mm 
\par
{\bf Acknowledgement.} We would like to thank M.C. Abbati for her interest on
 this work and useful suggestions.  
\par

\par
\end{document}